\documentclass[twocolumn,prd,nofootinbib]{revtex4}  
\usepackage{graphicx,amsmath,amssymb,xcolor}
\usepackage{subfigure}
\usepackage{color}

\newcommand{\beq}{\begin{eqnarray}}
\newcommand{\eeq}{\end{eqnarray}}

\begin{document}
\title{Unification of gauge and Yukawa couplings }
\author{Ammar Abdalgabar}
\email[Email: ]{aabdalgabar@gmail.com}
\affiliation{Department of Physics, Sudan University of Science and Technology, Khartoum 407, Sudan}
\author{Mohammed Omer Khojali}
\email[Email: ]{khogali11@gmail.com}
\affiliation{National Institute for Theoretical Physics; School of Physics, University of the Witwatersrand, Wits 2050, South Africa}
\author{Alan~S.~Cornell}
\email[Email: ]{alan.cornell@wits.ac.za}
\affiliation{National Institute for Theoretical Physics; School of Physics, University of the Witwatersrand, Wits 2050, South Africa}
\author{Giacomo Cacciapaglia}
\email[Email: ]{g.cacciapaglia@ipnl.in2p3.fr }
\affiliation{Universit\'e de Lyon, Universit\'e Lyon 1, CNRS/IN2P3, UMR5822 IPNL, F-69622 Villeurbanne, France}
\author{Aldo Deandrea}
\email[Email: ]{deandrea@ipnl.in2p3.fr}
\affiliation{Universit\'e de Lyon, Universit\'e Lyon 1, CNRS/IN2P3, UMR5822 IPNL, F-69622 Villeurbanne, France}
\begin{abstract}
The unification of gauge and top Yukawa couplings is an attractive feature of gauge-Higgs unification models in extra-dimensions. This feature is usually considered difficult to obtain based on simple group theory analyses. We reconsider a minimal toy model including the renormalisation group running at one loop. Our results show that the gauge couplings unify asymptotically at high energies, and that this may result from the presence of an UV fixed point. The Yukawa coupling in our toy model is enhanced at low energies, showing that a genuine unification of gauge and Yukawa couplings may be achieved.
\end{abstract}
\maketitle


The discovery of a Higgs boson at the LHC experiments opened a new era in particle physics. Aside from being the last missing particle predicted by the Standard Model (SM), it is allowing a direct probe of the electroweak (EW) symmetry breaking sector of the SM.
In particular, the fact that its mass is close to the EW scale itself, has materialised the issue of naturalness. 
Mass terms for scalar fields are not protected by any quantum symmetry, therefore any new physics sector that couples to it will feed into the value of the mass. In the SM, the EW scale seems to be shielded from high energy scales, like the Planck one, however, no reason for this is present in the SM itself. Another intriguing hint for new physics is the unification of gauge couplings, that occurs at high energies once one takes into account the renormalisation group evolution of the couplings. This has lead to the development of Grand Unified Theories (GUT). The fact that the mass of the top quark is close to the EW scale also suggests that the Yukawa coupling of the top may have a similar origin.

The emergence of low-scale extra-dimensions~\cite{Antoniadis:1990ew}, mainly supported by string theory constructions, opened new avenues for model building. One of the most interesting ideas is developed in Gauge-Higgs Unification (GHU) models~\cite{Hosotani:1983xw,Hatanaka:1998yp,Dvali:2001qr}. Extra-dimensional models, in fact, contain a special class of scalar fields, that arise as an additional polarisation of vector gauge fields aligned with the extra compact space. 
If the Higgs can be identified as such a scalar, its couplings with the fermions (the top quark in particular) are also related to the gauge coupling.
In addition, mass terms for the Higgs would be forbidden by gauge invariance in the bulk of the extra-dimensions. If the gauge symmetry is suitably broken by boundary conditions, a massless scalar emerges in the spectrum, whose potential is then radiatively generated and finite~\cite{Masiero:2001im,Antoniadis:2001cv}. 

The GHU models are rather attractive as they address, at the same time, gauge-Yukawa unification and naturalness. The main challenge is to find a gauge group, $\mathcal{G}_{\rm GHU}$, that successfully predicts the values of the SM couplings. The requirement that it contains the EW gauge symmetry of the SM, i.e. SU(2)$_L$ and the U(1)$_Y$ of hypercharge, and at the same time broken generators transforming as the Higgs doublet field, strongly limits the possible choices. Most of these possibilities, though, would seem to give incorrect predictions~\cite{Csaki:2002ur}.
In this letter we show that this conclusion is modified once the energy evolution of the couplings is properly taken into account. In fact, as the extra-dimensions are to be considered as an effective theory, the unified predictions are only valid at the cut-off of the theory. However, the experimental values refer to the EW scale, and the couplings may well change due to the running via renormalisation group equations. This fact is well studied and understood in extra-dimensional GUTs~\cite{Dienes:1998vh,Ghilencea:1998st}. Even though the cut-off of the effective theory may be rather low, the running in extra-dimensions is not logarithmic but follows a power law~\cite{Dienes:1998vh,Ghilencea:2003xy,Varin:2006wa}, thus it is much faster than in four dimensions. We will show that, taking into account the running, the tree-level predictions are strongly modified and the low energy values of the SM couplings can match the experimental values, even if starting from completely different tree-level values.
For the top Yukawa, the running tends to ease the tension due to the largeness of the top Yukawa at low energy compared to the gauge couplings.

\subsection*{Minimal SU(3) model with a bulk triplet}

We will focus here on the simplest GHU group that allows us to embed both the EW symmetry and the Higgs: $\mathcal{G}_{\rm GHU} =$ SU(3)$_W$~\cite{Scrucca:2003ra}.
This group, of rank 2 like the EW symmetry, contains an SU(2)$\times$U(1) subgroup that can be identified with the gauged EW one. Furthermore, the remaining 4 broken generators correspond to a doublet of SU(2) with non-vanishing hypercharge, like the Higgs doublet in the SM. Fixing the hypercharge of the doublet fixes the relation between the SU(2) and U(1) couplings.
Finally, a fermion field in the fundamental representation decomposes into a doublet and singlet of the SU(2): once the hypercharge of the Higgs candidate is fixed, the hypercharges of the doublet and singlet matches those of the left-handed quarks and the right-handed down-type ones. 
While we would like to describe the top quark as a bulk field, we will consider this simple model as a toy to test our idea.  
Note that other SM fermions can be added as localised degrees of freedom~\cite{Csaki:2002ur,Scrucca:2003ra}, however, their couplings to the bulk Higgs will be suppressed, thus explaining fermion masses below the EW scale.
The SU(3) predictions for the gauge and Yukawa couplings, in terms of the unified coupling $g_{\rm GHU}$, are shown in Table~\ref{tab:couplings} together with the SM values of the couplings at the EW scale (i.e. $M_Z$). 
For the Yukawa we consider the top Yukawa as our benchmark value because it is the largest one. 
It is clear that the tree-level GHU predictions are different from the SM values, however, they only apply at the cut-off of the effective theory, which may be very far from the EW scale. We show that the running will strongly modify the predictions.

 \begin{table}[tb]
 \begin{center}
 \begin{tabular}{l|c|c|c|c}
   & SU(2)$_L$ & U(1)$_Y$ & Yuk. & SU(3)$_c$\\
\phantom{\Big(}     & $g$  & $g'$ & $y$ & $g_s$ \\ 
\hline \hline
\phantom{\Big(}SU(3) GHU &  $g_{\rm GHU}$ & $\sqrt{3}\ g_{\rm GHU}$ & $g_{\rm GHU}/\sqrt{2}$ & -\\
\hline
\phantom{\Big(}SM & $0.66$ & $0.35$ & $1.0$ & $1.2$ \\
\end{tabular} 
\end{center}
\caption{Gauge and top Yukawa couplings in the SU(3) GHU model compared to the SM values at the $M_Z$ scale. We also include for completeness the QCD coupling.} \label{tab:couplings}
\end{table}

We thus study the running effects in a concrete model based on a single extra-dimension compactified on an interval $S^1/\mathbb{Z}_2$. The boundary conditions at the two end points of the interval, $x_5 = 0$ and $x_5 = \pi R$ (where $R$ is the radius of the extra-dimension), are such that the GHU group is broken to the EW one. The spectrum will thus contain massless gauge bosons plus a massless scalar associated to the broken generators. Furthermore, the bulk fermion transforming as the fundamental of SU(3)$_W$ is assigned boundary conditions such that only two massless fermions appear and we identify them with the third generation quark doublet and down-type singlet (the missing SM fermions are assumed to be localised). At low energy, therefore, the field content matches that of the SM. The running of the couplings will be affected by the presence of the Kaluza-Klein (KK) states once the mass thresholds are met, starting at $m_{\rm KK} = 1/R$.

\begin{figure}[h!]
\center
\includegraphics[scale=0.45]{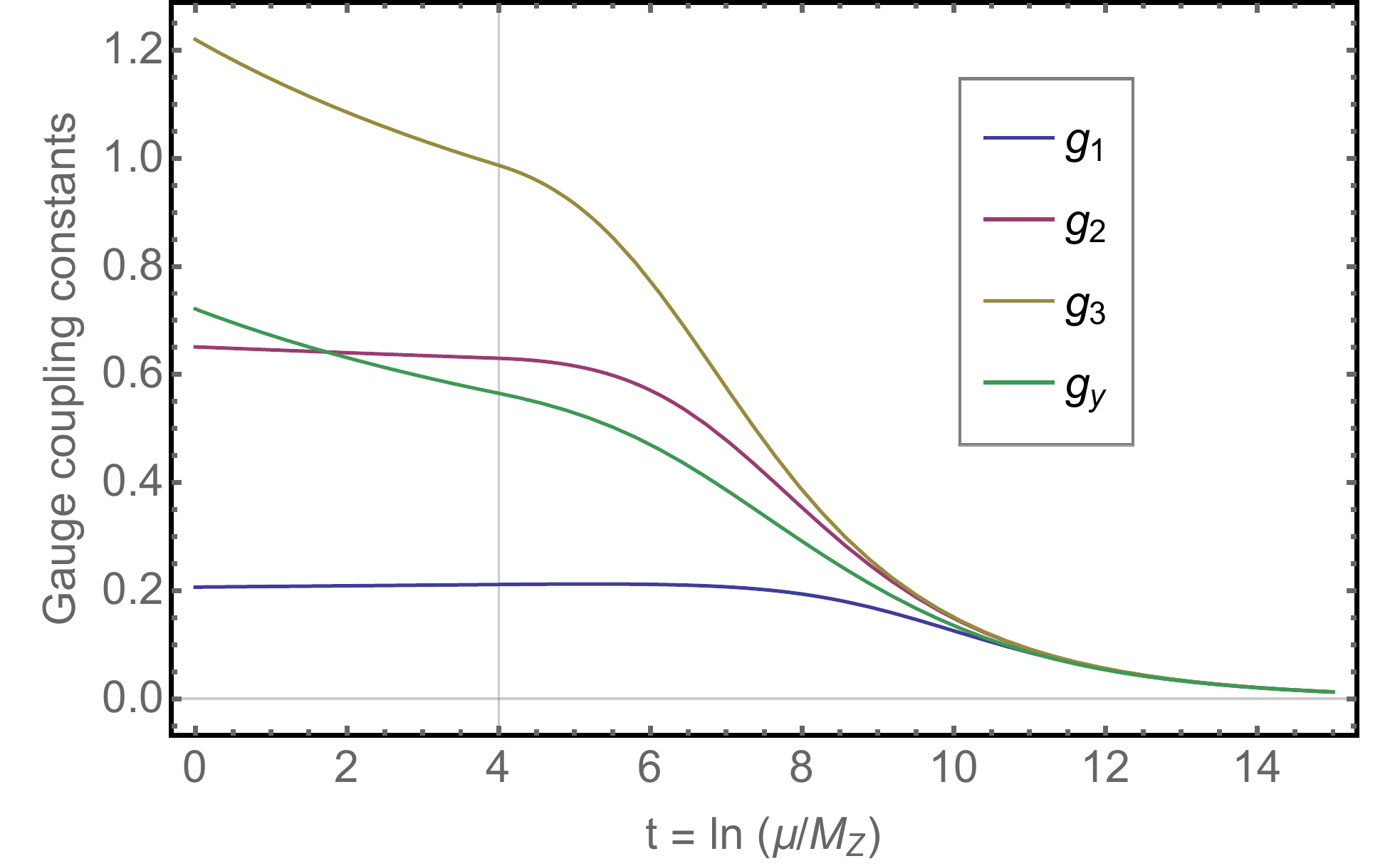}
\caption{Running of the normalised gauge and Yukawa couplings for the SU(3) GHU model, for $1/R = 5$ TeV. The first KK mode enters at $t_{\rm KK} \sim 4.0$.}
\label{fig:plot}
\end{figure}

\begin{figure}[h!]
\center
\includegraphics[scale=0.45]{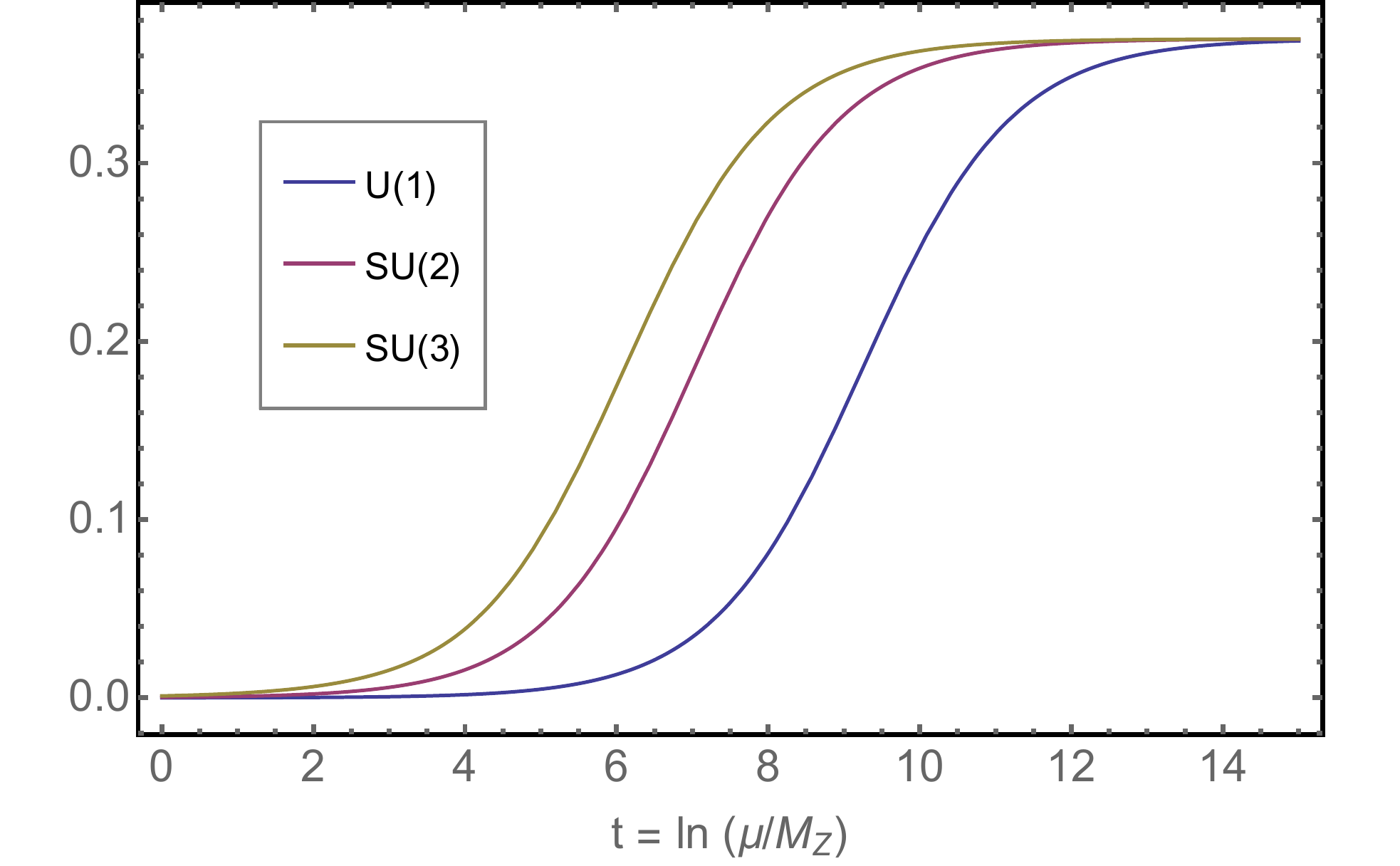}
\caption{5D NDA loop factor as a function of the energy, for $1/R = 5$ TeV.}
\label{fig:plot2}
\end{figure}

In Figure~\ref{fig:plot} we show the running of the couplings as a function of the energy scale $\mu$, normalised to the unified values as in Table~\ref{tab:couplings}:
\begin{equation}
\{g_1, g_2, g_3, g_y\} = \left\{ \frac{g'}{\sqrt{3}},  g,  g_s, \sqrt{2}\ y\right\}\,.
\end{equation}
The normalisations simply follow from the group theory structure of the SU(3)$_W$ matrices, while the QCD coupling is, in principle, unrelated.
The couplings follow SM evolutions up to the scale where the first KK resonances appear, i.e. 
\begin{equation}
t_{\rm KK} = \ln \frac{1}{M_Z R}\,.
\end{equation}
From there on the running is modified by the extra-dimensions, and it features the expected linear behaviour.
The figure clearly shows that the gauge couplings asymptotically tend to the same value. This is more evident from the plot in Figure~\ref{fig:plot2}, where we show, as a function of the energy, a naive estimate of the 5-dimensional loop factor, obtained by using naive dimensional analysis (NDA)~\cite{Weinberg:1978kz,Manohar:1983md}:
\begin{equation} \label{eq:NDA}
\alpha_i^{\rm NDA} (\mu) \sim \frac{g_i^2 (\mu)}{8 \pi} \mu R\,.
\end{equation}
While all the couplings run asymptotically to zero, their ratio clearly tends to 1. Thus it looks as if the unified value of the gauge couplings is an UV attractor of the one loop running. 
It may seem surprising that the strong coupling also falls very close. However, the GHU model contains two SU(3) gauge structures, one associated to QCD and the other to the EW gauge sector, and the bulk fermion is a bi-fundamental. This allows the existence of a $\mathbb{Z}_2$ symmetry between the two sectors at high energy that implies equal couplings.
Note, finally, that the NDA loop factor, which can be thought of as a 5D 't Hooft coupling (as $\mu R$ counts the number of KK tiers below energy $\mu$), can be used as a marker of the energy where the calculability of the extra-dimensional theory is lost.
The fact that the values stay small seems to suggest that the theory under study may have a more extended validity than previously thought.

The initial value of the Yukawa coupling, corresponding to $y (m_Z) = 0.51$, is tuned to achieve unification in the UV. 
This value depends only mildly on the scale of the extra-dimension $1/R$. It should be noted that the running of the Yukawa coupling does not follow the gauge ones at high energy, due to the fact that the compactification of the extra-dimension clearly singles out the scalar component of the bulk gauge field. However, in the UV, the running needs to be replaced by the running of the 5D gauge coupling.
Our results show that the value of the Yukawa coupling at low energy is larger than the values at unification, $y = g_2/\sqrt{2}$, however the enhancement is not enough to explain the Yukawa coupling of the top, $y=1$. It should be remarked that the value we obtain is a solid prediction of this toy model. Nevertheless two loop corrections, and the embedding of the top in a more realistic model, may further improve the agreement.

One possibility is to replace the bulk fermion triplet with a larger representation that can contain a singlet with the correct hypercharge to match the right-handed top: the minimal possibility is to use a 2-index symmetric representation (sextet). The sextet would contain a doublet and singlet matching the quantum numbers of the SM quarks, plus an SU(2) triplet. Thus, one can define two independent Yukawa couplings. Furthermore, the triplet acquires a mass by marrying to a localised chiral fermion, which is also needed to cancel residual 4-dimensional gauge anomalies. We also performed the running in this model, following the same prescriptions as before. However, we noticed that the NDA loop factor estimate for the EW gauge couplings run to non-perturbative values well before unification occurs, thus rendering  the perturbative running unreliable. This result seems to indicate that only models with small representations of the bulk gauge symmetries can provide useful predictions for the low energy values of the couplings in the model.

\subsection*{Details of the calculation}

The renormalisation group equations allow us to resum the leading energy-dependent corrections to any coupling in terms of a differential equation. The solutions are energy-dependent couplings whose values run with the scale at which the physics is probed. While in four dimensions the running is logarithmic, in five dimensional models it becomes linear in the energy.
The generic structure of the running of the gauge couplings at one loop level is given by~\cite{Bhattacharyya:2006ym} 
\begin{equation} \label{eq:runG}
 16\pi^{2}\ \frac{d g_{i}}{d t} = b^{\rm SM}_{i}\ g_{i}^{3} + (S(t)-1)\,b^{\rm GHU}_{i}\ g_{i}^{3}\,,
\end{equation}
where $t = \ln\left(\mu/M_{Z}\right)$ and contains the energy scale parameter $\mu$.
We chose to use the $Z$ mass as a reference scale, so that for $\mu = M_Z$ we have $t=0$ and we can fix the initial conditions of the running.
The coefficients $b_i^{\rm SM}$ and $b_i^{\rm GHU}$ can be computed once the field content of the model is specified: the former are equal to the values in the SM, while the latter include the effects of the KK modes in the bulk of the extra-dimension. This effect only starts contributing above the mass of the first mode, equal to the inverse radius $m_{\rm KK} = 1/R$.
The function $S(t)$, defined as 
\begin{equation}
S(t) = \left\{ \begin{array}{l}
\mu\,R = M_Z\, R\ e^t \qquad  \mbox{for} \quad \mu > 1/R\,, \\
1\qquad \mbox{for} \quad M_Z < \mu < 1/R, 
\end{array} \right.
\end{equation}
encodes the linear running due to the extra-dimension. 
This continuum approximation has been tested against the discrete sum over the KK modes, and the results are in excellent agreement.
For the SU(3) GHU model the $b$ coefficients for the SM gauge couplings, $g_i = \{g', g, g_s\}$, are
\begin{equation}
b^{SM}_{i} = \left[\frac{41}{10},\, -\frac{19}{6},\, -7\right], \quad b^{SU(3)}_{i} = \left[-\frac{17}{6},\, -\frac{17}{2},\, -\frac{17}{2}\right]\,.
\end{equation}
This result can be easily understood: $-17/2$ is the beta function of the unified SU(3) model (recall that $b_1^{SU(3)}$ has an additional normalisation of $1/3$), and the result matches the fact that each KK tier contains a complete representation of SU(3). For the hypercharge running the normalisation factor has been taken into account.

The asymptotic behaviour of the running of the gauge couplings can be easily understood when rewriting Eq.(\ref{eq:runG}) in terms of $\alpha^{\rm NDA}$ (as defined in Eq.(\ref{eq:NDA}))
\begin{equation} \label{eq:NDArun}
\frac{d \alpha^{\rm NDA}}{d t} = \alpha^{\rm NDA} + \frac{b^{SU(3)}}{\pi} (\alpha^{\rm NDA})^2\,,
\end{equation}
where we only retain the term proportional to $S(t)$ that grows with energy. As such, for negative $b$, the above equation allows for an UV fixed point, where the coupling stops running, that is
\begin{equation}
\left. \alpha^{\rm NDA}\right|_{\rm UV} = - \frac{\pi}{b^{SU(3)}} = \frac{2 \pi}{17}\,.
\end{equation}
The value above matches the numerical value we found in Figure~\ref{fig:plot2} and, as discussed earlier, it remains perturbative.
We also estimated the two loop contribution which adds to Eq.(\ref{eq:NDArun}) the following term
\begin{equation}
+ \frac{b^{SU(3)}_{\rm 2 loop}}{2 \pi^2} (\alpha^{\rm NDA})^3
\end{equation} 
with $b^{SU(3)}_{\rm 2 loop} = -44$. The zero of the beta function is marginally corrected and now appears at $\left. \alpha^{\rm NDA}\right|_{\rm UV} \sim 0.3$. This confirms that the perturbative expansion is well behaved.
The presence of an UV fixed point is less certain, as there are non-perturbative indications against its presence \cite{Gies:2003ic,Morris:2004mg}. 

Similarly, the general form of the running of the one loop $\beta$-function for the top Yukawa coupling $y_{t}$ can be written as~\cite{Cornell:2012qf}:
\begin{equation}
16 \pi^{2}\ \frac{d\,y_{t}}{d\,t} = \beta^{SM}_{t} +\Big(S(t)-1\Big)\, \beta^{\rm GHU}_{t},
\end{equation}
where
\begin{equation}
\beta_t = y_t  \left[ c_t\ y_t^2 + \sum_i d_i\ g_i^2 \right]\,.
\end{equation}
Computing the coefficients for the Yukawa running is not as straightforward as for the gauge ones: already at one loop, vertices involving different KK modes contribute. Thus to simplify the calculation, we assigned the SM values to the new couplings. Note though that the choice needs to be done in a consistent way. As such, we decided on the following policy: for couplings between bosons, we always associate the coupling to a gauge one, while couplings to fermions depend on the quantum numbers of the boson (thus for doublets we associate the coupling to the Yukawa)\footnote{Note that for larger bulk representations this is the only physically meaningful choice. For instance, in the case of a sextet, two Yukawa couplings can be identified that run differently from each other.}. We also checked that the numerical results do not depend crucially on this choice.
For the model under study the coefficients assume the following values
\begin{equation}
c_t^{\rm SM} = \frac{9}{2}\,, \quad c_t^{SU(3)} = \frac{21}{2}\,,
\end{equation}
and
\begin{equation}
d^{SM}_{i} = \left[-\frac{5}{12},\, -\frac{9}{4},\, -8\right], \quad d^{SU(3)}_{i} = \left[-\frac{35}{24},\, -\frac{39}{8},\, -4\right]\,.
\end{equation}
It is interesting to notice that imposing the unification relations between the EW couplings and the Yukawa, compare to Table~\ref{tab:couplings}, one would obtain a beta function of 
\begin{equation}
\frac{1}{2} c_t^{SU(3)} +3 d^{SU(3)}_{1} + d^{SU(3)}_{2} = -4\,,
\end{equation} 
which is the same value of $d^{SU(3)}_{3}$ for the QCD contribution. Thus, the running of the scalar coupling, even in the unification regime, is different from the running of the vector couplings. This is due to the intrinsic violation of 5D gauge invariance encoded in the compactification of the extra-dimension.

The evolution equations for the gauge couplings can be solved analytically as, at one loop level, they are not coupled. On the other hand, the Yukawa coupling is related to the gauge couplings, therefore we have performed a numerical calculation, whose results are given in Figure~\ref{fig:plot}.

\subsection*{Significance of our results}

Our results show that the running cannot be neglected and is crucial to test the feasibility of gauge-Higgs-Yukawa unification in extra-dimensions. We have performed a one loop calculation within the approximation of neglecting the finite parts of the loops. The result can be improved by including the finite contributions, that may also depend linearly on the energy~\cite{Contino:2001si} and thus be non-negligible. For increased accuracy the two loop running may also be computed. For the purpose of this letter, the accuracy we achieved at one loop is sufficient to enforce our conclusions.
The simplicity of this model contrasts previous attempts made in the literature to address the issue of the mismatch between tree-level predictions and the low energy SM values.
The value of the gauge couplings can be easily modified by adding an extra gauged U(1)$_X$ in the extra-dimension. The hypercharge is thus identified with a combination of the U(1) contained in the unified group $\mathcal{G}_{\rm GHU}$ and of the new U(1)$_X$, and the gauge coupling $g_X$ can be tuned to the correct value. Additionally, localised kinetic terms~\cite{Carena:2002me} for the SM gauge subgroups (that are not broken on the boundaries) also modify the unified relation.
The challenge presented by the top Yukawa is more critical.  One possibility is to embed the top in an higher dimensional representation in order to gain a group theory factor~\cite{Cacciapaglia:2005da} at the price of lowering the cut-off of the theory. Another possibility is to modify the geometry of the extra-dimension by including a curvature: in such a case, playing with the localisation of the zero mode wave functions, with an enhanced overlap with the Higgs being obtained. The latter mechanism has been used in warped space~\cite{Contino:2003ve,Hosotani:2005nz}, leading to a revival of composite Higgs models. 
Properly taking into account the running, maybe none of the above complications would be necessary.
Note that obtaining the masses of light fermions is rather easy, as one can use localisation in flat space to suppress the overlap with the Higgs~\cite{Grossman:1999ra}, or include light fermions as degrees of freedom localised on the boundaries~\cite{Scrucca:2003ra}.

\subsection*{Conclusions}

Running of couplings from the EW scale to the extra-dimension scale needs to be taken into account in order to obtain reliable results. When included, it allows us to obtain simple models of GHU where both the EW gauge couplings and the top Yukawa unify.
We have studied a toy model in five-dimensions, compactified on an interval $S^1/\mathbb{Z}_2$, with bulk gauge groups SU(3)$_c \times$ SU(3)$_W$ and a bulk fermion transforming as a bi-fundamental.
This simple structure is enough to describe the EW gauge sector unified in SU(3)$_W$. The fermions contained in the bulk fermion match a down-type quark, yet the effective Yukawa coupling is enhanced at low energies thanks to the running.
We show that the running allows us to match the value of the Weinberg angle at the EW scale, as well as larger than expected Yukawa couplings.
Unified values of the couplings appear as an attractor in the UV, providing an example of asymptotic unification. 
The QCD gauge coupling also unifies, suggesting that the double-SU(3) structure may be symmetric and may be embedded in a larger algebra.

\subsubsection*{Acknowledgements}
We acknowledge support from the CNRS PICS project no. 07552 and the Campus France PHC PROTEA 2017 project no. 38192RC. 
The work of AC and MK is also supported by the National Research Foundation of South Africa.
 GC and AD also acknowledge partial support from
the Labex-LIO (Lyon Institute of Origins) under grant ANR-10-LABX-66 and FRAMA (FR3127, F\'ed\'eration de Recherche ``Andr\'e
Marie Amp\`ere"). 
AD is partially supported by the ``Institut Universitaire de France''.



\begin{thebibliography}{99}


\bibitem{Antoniadis:1990ew}
  I.~Antoniadis,
  Phys.\ Lett.\ B {\bf 246} (1990) 377.


\bibitem{Hosotani:1983xw}
  Y.~Hosotani,
  Phys.\ Lett.\  {\bf 126B} (1983) 309.

\bibitem{Hatanaka:1998yp}
  H.~Hatanaka, T.~Inami and C.~S.~Lim,
  Mod.\ Phys.\ Lett.\ A {\bf 13} (1998) 2601
  [hep-th/9805067].

\bibitem{Dvali:2001qr}
  G.~R.~Dvali, S.~Randjbar-Daemi and R.~Tabbash,
  Phys.\ Rev.\ D {\bf 65} (2002) 064021
  [hep-ph/0102307].

\bibitem{Masiero:2001im}
  A.~Masiero, C.~A.~Scrucca, M.~Serone and L.~Silvestrini,
  Phys.\ Rev.\ Lett.\  {\bf 87} (2001) 251601
  [hep-ph/0107201].


\bibitem{Antoniadis:2001cv}
  I.~Antoniadis, K.~Benakli and M.~Quiros,
  New J.\ Phys.\  {\bf 3} (2001) 20
  [hep-th/0108005].



\bibitem{Csaki:2002ur} 
  C.~Csaki, C.~Grojean and H.~Murayama,
  Phys.\ Rev.\ D {\bf 67}, 085012 (2003)
  [hep-ph/0210133].

\bibitem{Dienes:1998vh}
  K.~R.~Dienes, E.~Dudas and T.~Gherghetta,
  Phys.\ Lett.\ B {\bf 436} (1998) 55
  [hep-ph/9803466].


\bibitem{Ghilencea:1998st}
  D.~Ghilencea and G.~G.~Ross,
  Phys.\ Lett.\ B {\bf 442} (1998) 165
  [hep-ph/9809217].


  
\bibitem{Ghilencea:2003xy}
  D.~M.~Ghilencea,
  Phys.\ Rev.\ D {\bf 70} (2004) 045011
  [hep-th/0311187].
  
\bibitem{Varin:2006wa}
  T.~Varin, J.~Welzel, A.~Deandrea and D.~Davesne,
  Phys.\ Rev.\ D {\bf 74} (2006) 121702
  [hep-ph/0610130].



\bibitem{Scrucca:2003ra} 
C.~A.~Scrucca, M.~Serone and L.~Silvestrini,
Nucl.\ Phys.\ B {\bf 669}, 128 (2003)
[hep-ph/0304220].


\bibitem{Weinberg:1978kz}
  S.~Weinberg,
  Physica A {\bf 96} (1979) 327.


\bibitem{Manohar:1983md}
  A.~Manohar and H.~Georgi,
  Nucl.\ Phys.\ B {\bf 234} (1984) 189.


\bibitem{Bhattacharyya:2006ym}
  G.~Bhattacharyya, A.~Datta, S.~K.~Majee and A.~Raychaudhuri,
  Nucl.\ Phys.\ B {\bf 760} (2007) 117
  [hep-ph/0608208].


\bibitem{Gies:2003ic}
  H.~Gies,
  Phys.\ Rev.\ D {\bf 68} (2003) 085015
  [hep-th/0305208].

\bibitem{Morris:2004mg}
  T.~R.~Morris,
  JHEP {\bf 0501} (2005) 002
  [hep-ph/0410142].


\bibitem{Cornell:2012qf} 
A.~S.~Cornell, A.~Deandrea, L.~X.~Liu and A.~Tarhini,
Mod.\ Phys.\ Lett.\ A {\bf 28}, no. 11, 1330007 (2013)
[arXiv:1209.6239 [hep-ph]].



\bibitem{Contino:2001si}
  R.~Contino, L.~Pilo, R.~Rattazzi and E.~Trincherini,
  Nucl.\ Phys.\ B {\bf 622} (2002) 227
  [hep-ph/0108102].

  
\bibitem{Carena:2002me}
  M.~Carena, T.~M.~P.~Tait and C.~E.~M.~Wagner,
  Acta Phys.\ Polon.\ B {\bf 33} (2002) 2355
  [hep-ph/0207056].
  
  
  \bibitem{Cacciapaglia:2005da} 
  G.~Cacciapaglia, C.~Csaki and S.~C.~Park,
  JHEP {\bf 0603}, 099 (2006)
  [hep-ph/0510366].

  
\bibitem{Contino:2003ve}
  R.~Contino, Y.~Nomura and A.~Pomarol,
  Nucl.\ Phys.\ B {\bf 671} (2003) 148
  [hep-ph/0306259].

  
\bibitem{Hosotani:2005nz}
  Y.~Hosotani and M.~Mabe,
  Phys.\ Lett.\ B {\bf 615} (2005) 257
  [hep-ph/0503020].
  
\bibitem{Grossman:1999ra}
  Y.~Grossman and M.~Neubert,
  Phys.\ Lett.\ B {\bf 474} (2000) 361
  [hep-ph/9912408].


  

\end{thebibliography}
\end{document}